\documentclass[12pt, a4paper]{article}

\usepackage{amsfonts}
\usepackage{amsmath}
\usepackage{amssymb}
\usepackage{ascmac}
\usepackage{dcolumn}
\usepackage{bm,here}
\usepackage{subfig}
\usepackage{comment}
\usepackage{ifpdf}
\usepackage{slashed}
\usepackage{colortbl}
\usepackage[dvipsnames]{xcolor}
\usepackage[mathscr]{eucal}
\usepackage{cite}
\usepackage[utopia]{mathdesign}

\usepackage[left = 20mm, right = 20mm, top = 25mm, bottom = 25mm]{geometry}

\usepackage{graphicx}     
\usepackage[bookmarksopen, 
colorlinks=true,linkcolor=dark_blue,citecolor=dark_red,urlcolor=dark_red,linktocpage=false]{hyperref}

\usepackage{multicol}
\definecolor{red}{rgb}{1,0,0}
\definecolor{blue}{rgb}{0,0,1}
\definecolor{orange}{rgb}{1,0.5,0}
\definecolor{dark_blue}{rgb}{0.0, 0., 0.6}
\definecolor{dark_red}{rgb}{0.7, 0., 0.}
\usepackage{framed}

\begin{document}

\newcommand{\rem}[1]{{ \textcolor{red}{[[$\spadesuit$\bf #1$\spadesuit$]]} }}

\renewcommand{\theequation}{\thesection.\arabic{equation}}
\renewcommand{\thefootnote}{$\natural$\arabic{footnote}}
\setcounter{footnote}{0}



\newcommand{\vev}[1]{ \left\langle {#1} \right\rangle }
\newcommand{\bra}[1]{ \langle {#1} | }
\newcommand{\ket}[1]{ | {#1} \rangle }
\newcommand{\EV}{ \ {\rm eV} }
\newcommand{\KEV}{ \ {\rm keV} }
\newcommand{\MEV}{\  {\rm MeV} }
\newcommand{\GEV}{\  {\rm GeV} }
\newcommand{\TEV}{\  {\rm TeV} }
\newcommand{\1}{\mbox{1}\hspace{-0.25em}\mbox{l}}
\newcommand{\Red}[1]{{\color{red} {#1}}}
\newcommand{\prn}[1]{\left( {#1} \right)}
\newcommand{\com}[1]{\left[ {#1} \right]}
\newcommand{\lmk}{\left(}  
\newcommand{\rmk}{\right)}
\newcommand{\lkk}{\left[}  
\newcommand{\rkk}{\right]}
\newcommand{\lhk}{\left \{ }  
\newcommand{\rhk}{\right \} }
\newcommand{\del}{\partial}  
\newcommand{\la}{\left\langle} 
\newcommand{\ra}{\right\rangle}
\newcommand{\half}{\frac{1}{2}}
\newcommand{\der}{\partial}  
\newcommand{\dd}{\mathrm{d}}
\newcommand{\Mpl}{M_{\rm Pl}}
\newcommand{\mg}{m_{3/2}}
\newcommand{\abs}[1]{\left\vert {#1} \right\vert}
\newcommand{\mphi}{m_\phi}
\newcommand{\Hz}{\ {\rm Hz}}
\newcommand{\Min}{{\rm Min}}
\newcommand{\Max}{{\rm Max}}
\newcommand{\Kahler}{K\"{a}hler }
\newcommand{\cphi}{\varphi}
\def\Mpl{M_{\rm pl}}


\begin{titlepage}

\begin{center}

\hfill UT-16-25\\
\hfill June, 2016\\

\vskip .75in

{\huge \bf 
Early Decay of Peccei-Quinn Fermion 
\\ \vspace{5mm} and the IceCube Neutrino Events
}

\vskip .75in

{\Large Yohei Ema,~ Takeo Moroi}

\vskip 0.25in

\begin{center}
{\em Department of Physics, Faculty of Science, 
\\ The University of Tokyo,  Bunkyo-ku, Tokyo 133-0033, Japan}
\end{center}

\end{center}
\vskip .5in

\begin{abstract}
IceCube observed high-energy neutrino flux in the energy region from TeV to PeV. 
The decay of a massive long-lived particle in the early universe can be the origin of the 
IceCube neutrino events, which we call an ``early decay scenario.''
In this paper, we construct a particle physics model that contains 
such a massive long-lived particle based on the Peccei-Quinn model.
We calculate the present neutrino flux, taking account of 
realistic initial energy distributions of particles produced by the decay
of the massive long-lived particle. 
We show that the early decay scenario naturally fits into the Peccei-Quinn model, 
and that the neutrino flux observed by IceCube can be explained in such a framework.
We also see that, based on that model, a consistent cosmological history that explains the abundance of 
the massive long-lived particle is realized.
\end{abstract}

\end{titlepage}

\tableofcontents

\renewcommand{\thepage}{\arabic{page}}
\setcounter{page}{1}

\newpage

\section{Introduction}
\label{sec:intro}
\setcounter{equation}{0}

After the observations of high energy cosmic ray neutrino events at
the IceCube~\cite{Aartsen:2013bka,Aartsen:2013jdh,Aartsen:2014gkd, Aartsen:2015zva, Aartsen:2016xlq}, 
the understanding of the origin of such high energy
neutrinos has become an important task in the field of astrophysics
and particle cosmology.  This is because, for the energy range of
$\mathcal{O}(10\,{\rm TeV})\lesssim 
E_\nu\lesssim \mathcal{O}(1\,{\rm PeV})$, the observed
flux is significantly larger than that expected from the background,
\textit{i.e.}, the atmospheric neutrinos.  There have been many attempts to understand the origin from
astrophysical~\cite{Kistler:2013my,Kalashev:2013vba,Murase:2013ffa,Laha:2013eev,Murase:2013rfa,Anchordoqui:2013qsi,Winter:2013cla,Razzaque:2013uoa,Ahlers:2013xia,Liu:2013wia,Lunardini:2013gva,Murase:2014foa,Tamborra:2014xia,Tjus:2014dna,Chang:2014hua,Dermer:2014vaa,Guo:2014laa,Murase:2015xka}\footnote{
	For more on this respect, see, \textit{e.g.} Refs.~\cite{Anchordoqui:2013dnh,Murase:2014tsa} and references therein.
} and particle physics~\cite{Feldstein:2013kka,Barger:2013pla,Esmaili:2013gha,Bai:2013nga,Bhattacharya:2014vwa,Higaki:2014dwa,Rott:2014kfa,Esmaili:2014rma,Murase:2015gea,Roland:2015yoa,Boucenna:2015tra,Ko:2015nma,Zavala:2014dla,Bhattacharya:2014yha,Fong:2014bsa,Kopp:2015bfa} points of
view, although no scenario is considered to be conclusive yet.

As one of the possibilities, it was pointed out that a long-lived
particle, with its mass being much higher than the PeV scale and its
lifetime shorter than the present age of the universe, may be the
origin of flux of high energy cosmic ray neutrinos \cite{Ema:2013nda}.\footnote{
	See Ref.~\cite{Anchordoqui:2015lqa} as a similar scenario.
}
(Hereafter, such a scenario is called as the ``early decay
scenario.'')  In the early decay scenario, even though the neutrino
produced by the decay of the long-lived particle has much larger
energy than the PeV scale, which is the maximum energy scale of the
deposited energy observed by IceCube, the energy is red-shifted so
that the present energy can be 
$\mathcal{O}(10\,{\rm TeV}) \lesssim E_\nu\lesssim
\mathcal{O}(1\,{\rm PeV})$.  
In Refs.~\cite{Ema:2013nda, Ema:2014ufa}, it was discussed that the decay of
a long-lived particle with its mass of $\mathcal{O}(10^{10}\,{\rm GeV})$ may
well explain the IceCube events, without specifying the
particle-physics model for the long-lived particle.

In considering the early decay scenario, one of the questions is the
origin of the mass scale of the long-lived particle.  In particular,
the IceCube events may suggest the existence of a new physics at such
a scale, into which the long-lived particle is embedded.  The
suggested mass scale is, as we mentioned above, 
$\sim 10^{10}\,{\rm GeV}$; one of the well-motivated new physics at such a scale is the
Peccei-Quinn (PQ) symmetry~\cite{Peccei:1977hh, Peccei:1977ur} 
as a solution to the strong CP
problem.  The suggested scale of the PQ symmetry breaking is
$10^{9}\,\text{GeV} \lesssim f_a \lesssim 10^{12}\,\text{GeV}$
[see, \textit{e.g.}, Refs.~\cite{Peccei:2006as,Kim:2008hd,Kawasaki:2013ae}],
which well agrees with the mass
scale required to realize the early decay scenario.  Importantly, in
hadronic axion models~\cite{Kim:1979if, Shifman:1979if}, 
the PQ mechanism requires the
existence of new colored fermions. Then, if the model is
embedded into grand unified theories (GUTs), there exist GUT partners
of the new colored fermions.  Some of these fermions (which we call PQ
fermions) are stable if they do not have any mixing with the standard
model (SM) fermions.  With the mixing being suppressed enough, those
fermions have long-enough lifetime required from the early decay
scenario.  As we will see below, some of the PQ fermion may play the
role of the long-lived particle whose decay explains the IceCube events.

In this paper, we study the possibility of explaining the flux of
cosmic ray high energy neutrinos observed by the IceCube using the
decay of the PQ fermion.  We assume that a color- and charge-neutral
fermion (called $X$) embedded into the PQ fermions becomes long-lived,
and that the decay of $X$ becomes the dominant source of the cosmic
ray neutrinos in the energy range of 
$\mathcal{O}(10\,{\rm TeV})\lesssim
E_\nu\lesssim \mathcal{O}(1\,{\rm PeV})$.  We concentrate on the case where the
lifetime of $X$ is much shorter than the present age of the universe,
and calculate the present flux of the neutrinos originating from the
decay of $X$.  For an accurate calculation of the present neutrino
flux, we carefully study the spectrum of the secondary neutrinos
produced from the decays of leptons and hadrons emitted from $X$.  Then, we
derive the mass, lifetime, and the relic abundance of the long-lived
particle to explain the observed neutrino flux.  We also discuss the
production mechanism of the right amount of the long-lived particle in
the early universe.  We will see that the neutral PQ fermion can be a
good candidate of the long-lived particle to realize the early decay
scenario, if its mass is $\mathcal{O}(10^{10}\,{\rm GeV})$ and also if its lifetime is
$\mathcal{O}(10^{11}\,{\rm sec})$.

The organization of this paper is as follows. 
In Sec.~\ref{sec:flux}, we specify the particle physics model we consider
in this paper, and calculate the present flux of neutrino produced by the decay of $X$
based on that model. 
In Sec.~\ref{sec:pheno}, we discuss various phenomenological aspects of our scenario.
In particular, as we will see in Sec.~\ref{subsec:present_flux}, the required abundance of $X$ to explain 
the IceCube events is quite tiny, and hence we discuss 
how such a tiny amount of $X$ is produced in the early universe.
Other cosmological concerns such as the dark matter 
and the fate of massive particles other than $X$ are also discussed there.
The last section is devoted to the summary.

\section{Present flux of neutrino}
\label{sec:flux}
\setcounter{equation}{0}

\subsection{Model}
\label{subsec:model}
Here we specify our particle physics model.
We introduce a heavy vector-like fermion $X_2$ that has the same Standard Model (SM) 
gauge charges as the SM lepton doublet. 
We take the Peccei-Quinn (PQ) model~\cite{Peccei:1977hh, Peccei:1977ur}
which solves the strong CP problem [see, \textit{e.g.}, Ref.~\cite{Kim:1986ax}]
as an example containing such a fermion.\footnote{
The gauge-mediated supersymmetry breaking model~\cite{Dine:1993yw, Dine:1994vc, Dine:1995ag}
is another candidate which also includes such a vector-like fermion, but we do not discuss it
in this paper.
}
In the hadronic axion model~\cite{Kim:1979if, Shifman:1979if}, an additional heavy quark
that carries the $\mathrm{U}(1)_\mathrm{PQ}$ charge is introduced. 
If the SM gauge groups are embedded in
the grand unified theory (GUT) such as SU(5), there must be an additional SU(2) doublet
fermion in the theory. After the PQ symmetry breaking, it acquires a mass 
of the order of the PQ scale.
We summarize the particle contents of our model in Table~\ref{tab:part_cont}.

\begin{table}[t]
  \begin{center}
  \begin{tabular}{l || ccc | cc}
	~~ & $\text{SU}(3)_C$ & $\text{SU}(2)_L$ & $\text{U}(1)_Y$ 
	& $\text{U}(1)_\text{PQ}$ & $\text{U}(1)_X$ \\ \hline
	$X_{3}$ & $\bar{{\bf 3}}$ & {\bf 1} & $+1/3$ & $+1$ & $+1$\\ 
	$X_{2}$ & ${\bf 1}$ & ${\bf 2}$ & $-1/2$ & $+1$ & $+1$ \\ 
	$\bar{X}_{3}$ & ${\bf 3}$ & ${\bf 1}$& $-1/3$ & $+1$ & $-1$ \\
	$\bar{X}_{2}$ & ${\bf 1}$ & ${\bf 2}$ & $+1/2$ & $+1$ & $-1$
  \end{tabular}
  \caption{\small Particle contents and their representation 
  		under the SM gauge group 
		as well as $\text{U}(1)_\text{PQ}$ and $\text{U}(1)_X$ in our model.
  		They are embedded into ${\bf 5}$ and ${\bf \bar{5}}$
		representations of the SU(5) GUT theory.
		Here, they are written as left-handed Weyl spinors.
		}
  \label{tab:part_cont}
  \end{center}
\end{table}

We identify the neutral component of $X_2$ as a massive long-lived particle (which we call $X$).
The long lifetime can be naturally realized as follows.
We can assign a conserved $\mathrm{U}(1)_X$ charge to the PQ fermions if there is no mixing term
with the SM sector, and hence the lightest fermion in the PQ sector is stable in that case.
In general, the colored partner of $X_2$ is heavier
than $X_2$ because of the renormalization group (RG) running effect
from the GUT scale to the PQ scale.
The charged component of $X_2$ also acquires mass from radiative corrections
after the electroweak phase transition as we will see in Sec.~\ref{subsec:lifetime}.
Thus, $X$ is typically the lightest fermion in the PQ sector,
and is stable if there is no mixing term between the SM and PQ sectors.
If the $\mathrm{U}(1)_X$ is (slightly) broken, the following Yukawa term is allowed
as an interaction between the PQ-sector and the SM-sector:\footnote{
A mass mixing term is equivalent to the Yukawa term after 
diagonalizing the mass matrix.
}
\begin{align}
\mathcal{L}_{\rm mix} = \epsilon_{y} \bar{X}_{2L}He_{R} + \mathrm{h.c.},
\label{eq:mixing}
\end{align}
where $H$ and $e_{R}$ are the SM Higgs doublet and right-handed lepton, 
$X_{2L}$ is the left-handed component of $X_{2}$, and we suppress the flavor indices for simplicity.
It is technically natural in 't Hooft's sense~\cite{'tHooft:1979bh} to take the coupling $\epsilon_y$
small because $\mathrm{U}(1)_X$ symmetry is restored in the limit of $\epsilon_y\rightarrow 0$.
The decay rate of $X$ is suppressed by $\epsilon_y$,
and $X$ can be a massive long-lived particle.

Once the Yukawa term is fixed, $X$ decays as $X \rightarrow l^{\pm} W^{\mp}$ 
after the electroweak symmetry breaking, 
where $l^\pm$ is the SM charged lepton and $W^\pm$ is the $W$ gauge boson.
The decay rate  $\Gamma_X$ is given by
\begin{align}
\tau_X^{-1} \equiv \Gamma_{X} \sim \alpha\epsilon_y^2 m_X,
\end{align}
where $\tau_X$ is the lifetime of $X$ and 
$\alpha$ is a numerical constant of order $\mathcal{O}(10^{-2})$.
For simplicity, we assume that $X$ equally couples to electron, muon
and tau lepton. Also we assume that the CP symmetry is conserved in the $X$ decay so that
particles and anti-particles are equally produced.
As we will see in Sec.~\ref{subsec:present_flux}, this decay mode is enough to explain the 
high-energy neutrino events observed by IceCube.

\subsection{Boltzmann equation}
\label{subsec:boltzmann}
In this subsection, we explain the time evolution of the high-energy neutrinos produced
by the decay of the massive long-lived particle $X$ in the early universe.
We follow the calculation procedure given in Ref.~\cite{Ema:2014ufa}.

Neutrinos propagate after produced by the decay of $X$, being sometimes scattered by
the background particles (especially neutrinos). In order to obtain the present flux of neutrino, we numerically
solve the following Boltzmann equation:
\begin{align}
  \left(\frac{\partial}{\partial t} + 2 H
    - H E \frac{\partial}{\partial E} \right) \Phi_{\nu, l}(t, E) 	
  = &
  -\gamma_{\nu, l}(t;E) \Phi_{\nu, l}(t, E) 
  \nonumber \\ & +
  \int dE^{\prime} 
  \Phi_{\nu, n}(t, E^{\prime})
  \frac{d \gamma_{\nu, nm}(t;E^{\prime}, E)}{d E}
  P_{ml}(t,E) \nonumber \\ &
  + S_{\nu, m}(t, E) P_{ml}(t,E),
  \label{eq:Boltzmann}
\end{align}
where $t$ is the cosmic time, $H$ is the Hubble parameter, 
$\Phi_{\nu, l}(t, E)$ is the neutrino flux in $l$-th flavor, $S_{\nu, l}(t, E)$ is the source term, 
$\gamma_{\nu, l}(t;E)$ is the scattering rate
and $d\gamma_{\nu, nm}(t; E^{\prime}, E)/dE$ is the differential neutrino production rate with $E^{\prime}$ ($E$) being 
the energy of the initial-state (final-state) neutrino. 
The Roman indices denote the flavors of the neutrinos.
The flux is related to the number density of neutrino as 
$n_{\nu, l}(t) = 4\pi\int dE\, \Phi_{\nu, l}(t, E)$ where $n_{\nu, l}(t)$ is the number density of
neutrino in $l$-th flavor. Moreover, we have introduced the ``transition probability" $P_{ml}(t, E)$ to take into account 
the neutrino oscillation effects. For more details, see Ref.~\cite{Ema:2014ufa}.

From now, we explain more details about the treatment of the source term.
Since we assume that the source of the high-energy neutrinos is the decay of $X$, the source term is given by
\begin{align}
S_{\nu, l}(t, E) &= \frac{1}{4\pi}\frac{n_{X}(t)}{\tau_{X}} \frac{dN_{\nu, l}^{(X)}}{dE} \nonumber \\
&= \frac{1}{4\pi}\frac{Y_{X}}{\tau_{X}}s(t)e^{-t/\tau_{X}}\frac{dN_{\nu, l}^{(X)}}{dE},
\end{align}
where $dN_{\nu, l}^{(X)}/dE$ is the energy distribution of neutrinos produced by the decay of one $X$,
$\tau_{X}$ is the lifetime of $X$, $n_{X}$ is the number density of $X$ and $Y_{X}$ is the yield variable of $X$
which is defined as
\begin{align}
Y_{X} \equiv \left[\frac{n_{X}(t)}{s(t)}\right]_{t\ll \tau_{X}},
\end{align}
with $s$ being the entropy density.
The energy distribution has two contributions:
\begin{align}
	\frac{d N_{\nu, l}^{(X)}}{dE} = \frac{d N_{\nu, l}^{(X, \text{dir})}}{dE} 
	+ \frac{d N_{\nu, l}^{(X, \gamma\gamma)}}{dE}.
\end{align}
Neutrinos are directly produced from the decay of $X$, and 
the contribution from such neutrinos is expressed as
$dN_{\nu, l}^{(X, \text{dir})}/dE$.
On the other hand, the decay of $X$ also produces electromagnetic particles, especially 
photons, electrons and positrons. These electromagnetic particles produce
high-energy neutrinos by the scattering with 
the cosmic microwave background (CMB) photons,
and we denote such a contribution to the flux 
as $dN_{\nu, l}^{(X, \gamma\gamma)}/dE$.
We estimate
\begin{align}
	\frac{dN_{\nu, l}^{(X, \gamma\gamma)}}{dE} 
	= \int d\epsilon \prn{\frac{dN_{\gamma}^{(X)}}{d\epsilon} + \frac{dN_{e^{\pm}}^{(X)}}{d\epsilon} }
	\frac{dN_{\nu, l}^{(\gamma\gamma)}(\epsilon, E)}{dE},
\end{align}
where $dN_{\gamma}^{(X)}/d\epsilon$ and $dN_{e^\pm}^{(X)}/d\epsilon$ are energy distributions of
photon and $e^\pm$ produced by the decay of one $X$, and
$dN_{\nu, l}^{(\gamma\gamma)}/dE$ is the spectrum of neutrinos produced by the double-photon
pair creation processes with CMB photons. 
Here $e^{\pm}$ is converted to photons with (almost) identical energy via the inverse
Compton scattering process.
We have used {\tt PYTHIA}
package~\cite{Sjostrand:2006za, Sjostrand:2007gs} 
to calculate $dN_{\nu,l}^{(X)}/dE$,
with the decay mode determined by Eq.~\eqref{eq:mixing}.
For more details on the treatment of high-energy electromagnetic particles,
see Ref.~\cite{Ema:2014ufa}.

Once the $X$'s mass $m_{X}$ is fixed, we can calculate 
the energy distribution $dN_{\nu, l}^{(X)}/dE$.
Thus, the source term, and also the present flux of neutrino are determined by the following 
three parameters:
\begin{align}
m_{X}, ~~ z_{*} \equiv z(\tau_{X}), ~~ Y_{X}.
\end{align}
We use the redshift $z_{*}$, at which the cosmic time is equal to $\tau_{X}$,
to parametrize $\tau_{X}$.
It depends on the Yukawa coupling $\epsilon_y$.

\subsection{Present flux and IceCube results}
\label{subsec:present_flux}
From now, we discuss implications of the early-decay scenario for
the high-energy neutrino events observed by IceCube.

The IceCube collaboration observed 
high-energy neutrinos~\cite{Aartsen:2013bka,Aartsen:2013jdh,Aartsen:2014gkd, Aartsen:2015zva,  Aartsen:2016xlq}.
In the three-year observation, they detected 37 events in the energy region of 
$30\,\text{TeV} \lesssim E \lesssim 2\,\text{PeV}$, three of which are with the deposited energy
of $1\,\text{PeV} \lesssim E \lesssim 2\,\text{PeV}$. 
Given that the expected backgrounds are $8.4\pm 4.2$ events for the atmospheric muon
and $6.6^{+5.9}_{-1.6}$ events for the atmospheric neutrinos,
it indicates a new source of the high-energy neutrinos.
In particular, in Ref.~\cite{Aartsen:2015knd}, 
they performed a detailed analysis of the high-energy neutrino flux.\footnote{
	See also Refs.~\cite{Chen:2013dza,Chen:2014gxa,Watanabe:2014qua,Palomares-Ruiz:2015mka,Vincent:2016nut}.
} Assuming the power-law,
the best fit all-flavor flux\footnote{
	Assuming the flavor universality, we obtain the per-flavor flux
	by dividing it by three.
}
for the energy region between
$25\,\text{TeV}$ and $2.8\,\text{PeV}$ is given by\footnote{
	A power-law flux with a spectral index $\gamma = 2$ and an exponential cut-off
	might be disfavored compared to a free spectral index one.
}
\begin{align}
	E^2\sum_{l = e, \mu, \tau}\Phi_{\nu, l}(E) 
	= \prn{6.7^{+1.1}_{-1.2}}\times 10^{-8} \prn{\frac{E}{100\,\text{TeV}}}^{-0.50\pm0.09}
	\text{GeV}\,\text{cm}^{-2}\,\text{s}^{-1}\,\text{sr}^{-1}.
\end{align}
In addition, in Ref.~\cite{Aartsen:2015zva}, the IceCube collaboration
shows their preliminary results of four-year observation of
high-energy neutrinos.  They detected additional 17 events in the
energy region of $30\,\text{TeV} \lesssim E \lesssim 400\,\text{PeV}$
in the fourth year.  The best fit value of the spectral index for all
four-year events with $60\,\text{TeV} < E < 3\,\text{PeV}$ is given by
$\gamma = -2.58 \pm 0.25$, and hence the expected flux might be even
softer after taking into account the fourth-year results.  We also
note that, recently, the highest energy charged current muon neutrino
event with the reconstructed muon energy of $4.5\pm 1.2\ {\rm PeV}$
was reported \cite{Aartsen:2016xlq}.

\begin{figure}[t]
\begin{minipage}{0.5\hsize}
\begin{center}
\includegraphics[scale = 0.8]{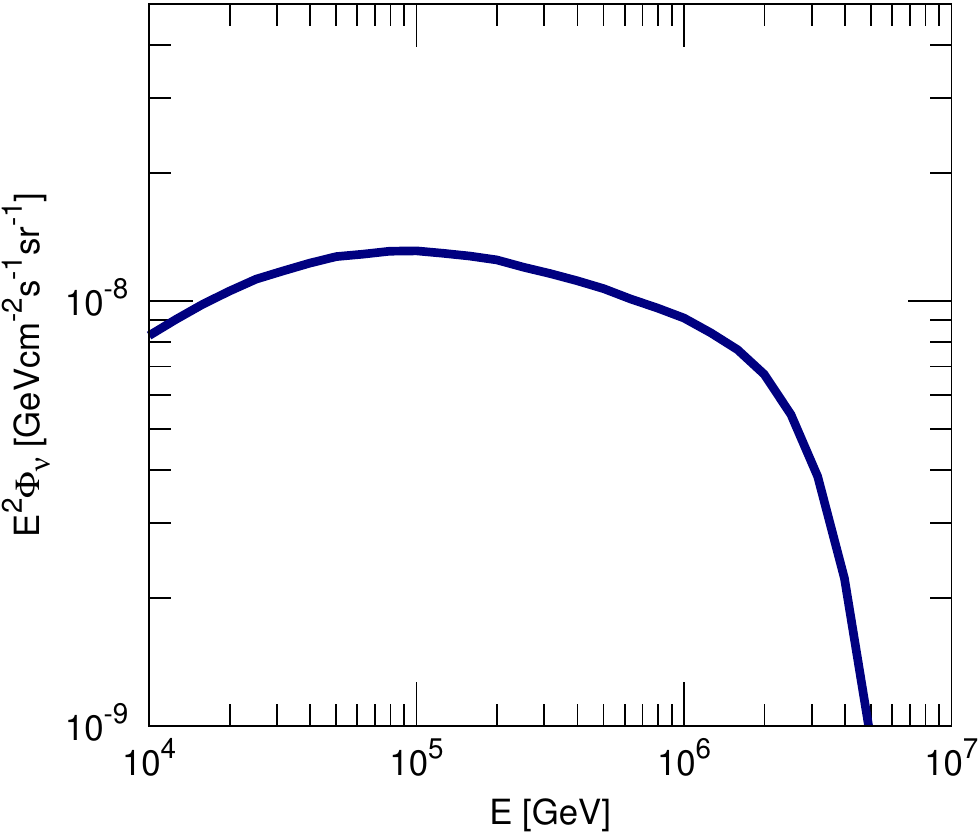}
\end{center}
\end{minipage}
\begin{minipage}{0.5\hsize}
\begin{center}
\includegraphics[scale = 0.8]{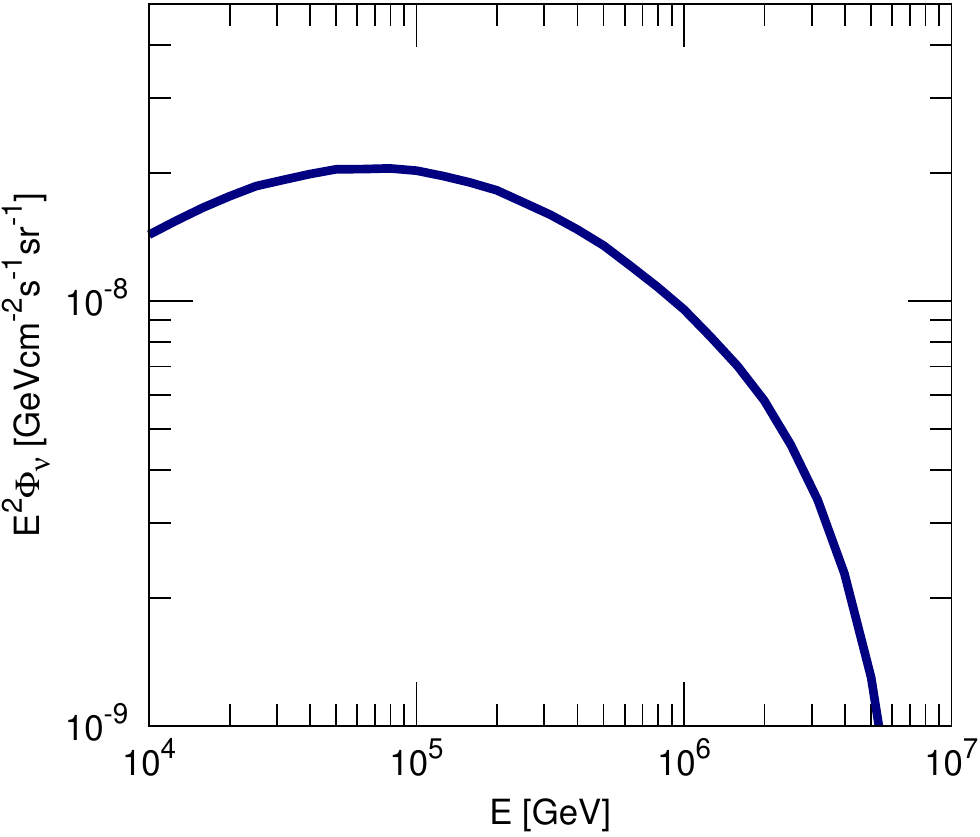}
\end{center}
\end{minipage}
\caption { \small
	The present per-flavor neutrino flux for two sample points. 
	Left panel: $(m_X,~ 1+z_*,~ Y_X) = (4\times 10^{10}\,\text{GeV},~ 8\times 10^{3},~ 8\times 10^{-27})$.
	Right panel: $(m_X,~ 1+z_*,~ Y_X) = (6\times 10^{10}\,\text{GeV},~ 1\times 10^{4},~ 1\times 10^{-26})$.
	The branching ratio is $\text{Br}\,(X \rightarrow l^{\pm} W^{\mp}) = 1$.
	We take the produced leptons as flavor-universal,
	and also assume that particles and anti-particles are equally produced.
	We plot the flavor-averaged value of the neutrino flux.
	The present neutrino flux is almost flavor-universal after taking account of
	the neutrino scattering and the neutrino oscillation.
}
\label{fig:flux}
\end{figure}

In order to show that the early-decay scenario can explain the IceCube neutrino events, 
we take the following two sample points in our analysis:
\begin{itemize}

\item point~1 [left panel]:
$(m_X,~ 1+z_*,~ Y_X) = (4\times 10^{10}\,\text{GeV},~ 8\times 10^{3},~ 8\times 10^{-27})$.

\item point~2 [right panel]:
$(m_X,~ 1+z_*,~ Y_X) = (6\times 10^{10}\,\text{GeV},~ 1\times 10^{4},~ 1\times 10^{-26})$.

\end{itemize}
We show the present neutrino per-flavor flux for these two sample
points in Fig.~\ref{fig:flux}.  The left panel corresponds to the
point~1, and the right panel does to the point~2.  The flux is close
to the $E^{-2}$ power-law with an exponential cut-off in the case of
the point~1, while the spectral index of the flux is rather close to
$\gamma \simeq -2.5$ in the case of the point~2.  Both cases give the
spectral index currently consistent with the IceCube results.  We
emphasize here that the spectral index is quite sensitive to $m_X$ and
$z_*$ if $z_* \sim \mathcal{O}(10^3)$ and $m_X \sim
\mathcal{O}(10^{10}\,\text{GeV})$.  This is because the neutrino
scattering effects begin to be effective at around such an epoch.
An ${\cal O}(10\%)$ tuning of the parameters is necessary to
obtain the neutrino flux with the spectral index in the range,
\textit{e.g.}, between $-2$ and $-2.5$.  In addition, the highest
energy muon neutrino event may be also due to the neutrinos produced
in the early decay scenario if the neutrino spectrum extends up to
$3\mathchar`-4\ {\rm PeV}$ (which is the case with the above two sample points),
assuming that a significant fraction of the energy of the primary muon
neutrino is carried away by the muon;
using the total exposure of the IceCube detector
given in Ref.~\cite{Aartsen:2016xlq}, the expected number of events
for $2.5 < E < 4~\mathrm{PeV}$ is 
$0.5$ for both the points~1 and~2.

We also note here that, for the energy of $E\simeq 6.3\ {\rm
    PeV}$, the event rate of the anti-electron neutrino is enhanced
  because of the on-pole exchange of the $W$-boson (so-called the
  ``Glashow resonance''~\cite{Glashow:1960zz}), which gives a
  stringent bound on the neutrino flux.  With four years of
  observation~\cite{Aartsen:2015zva}, no event was reported with the
  energy deposit of $E\simeq 6.3\ {\rm PeV}$.  Using the effective
  area provided by the IceCube collaboration,\footnote{
   For the data, see
    \href{https://icecube.wisc.edu/science/data/HE-nu-2010-2012}
    {https://icecube.wisc.edu/science/data/HE-nu-2010-2012}.
  } the expected number of events at the Glashow resonance is
  calculated to be $0.05$ and $0.1$ for the points 1 and 2,
  respectively, for 1347 days of livetime (which is the livetime of
  the four years of data).  Thus, in the early decay scenario, the
  neutrino flux at the Glashow resonance can be suppressed enough due
  to the exponential suppression of the the neutrino flux above a few
  PeV. Future study of neutrino events at the Glashow resonance
will be helpful to further constrain the spectral shape of the
neutrino flux.

The neutrino scattering with the background neutrinos produces 
electromagnetic particles, and such electromagnetic particles
may slightly modify the thermal history of the universe, leaving the trail 
in the present universe. In the case of our interest, constraints come from
the CMB $y$- and $\mu$-type distortions~\cite{Zeldovich:1969ff,Sunyaev:1970er,Danese:1982,Burigana:1991,Hu:1992dc,Burigana:1995,Chluba:2011hw} and
the big bang nucleosynthesis (BBN)~\cite{Kawasaki:2004yh,Kawasaki:2004qu}.
We estimate the $y$ and $\mu$ parameters as
$(y,~ \mu) = (4\times 10^{-9},~ 2\times 10^{-10})$ for the point~1
and $(y,~ \mu) = (6\times 10^{-9},~ 5\times 10^{-10})$ for the point~2.
Although they are consistent with the current the CMB observations~\cite{Mather:1993ij,Fixsen:1996nj},
future experiments such as 
PIXIE~\cite{Kogut:2011xw} and PRISM~\cite{Andre:2013nfa}
might have a chance to detect the CMB distortions caused by the decay of $X$.
We have also checked that the points 1 and 2 are not excluded 
by the BBN constraints~\cite{Kawasaki:2004yh,Kawasaki:2004qu}.

There are three comments. First, we comment on the angular distribution of the neutrino flux.
In the early-decay scenario, the neutrino flux is isotropic, and hence it is currently 
consistent with the IceCube results. It is in contrast to the case of 
the dark matter (DM) decay scenario~\cite{Feldstein:2013kka,Barger:2013pla,Esmaili:2013gha,Bai:2013nga,Bhattacharya:2014vwa,Higaki:2014dwa,Rott:2014kfa,Esmaili:2014rma,Murase:2015gea,Roland:2015yoa,Boucenna:2015tra,Ko:2015nma}.
In the DM decay scenario, the IceCube events are assumed to be
explained by the decay of DM, and hence 
the galactic contribution dominates over the extragalactic one.
Thus, a large fraction of neutrino comes from the galactic center region,
resulting in a sizable dipole asymmetry~\cite{Ema:2013nda}.
Further study on the angular distribution may be helpful to distinguish these two scenarios.

Second, we comment on the flavor composition of the neutrino flux.
So far the flavor ratio of the neutrino flux observed by IceCube is consistent with
the universal one $\Phi_{\nu, e}:\Phi_{\nu, \mu}:\Phi_{\nu, \tau} = 1:1:1$~\cite{Aartsen:2015knd,Aartsen:2015ivb}.
In the early-decay scenario, the present neutrino flux 
is almost flavor-universal after taking into account 
the neutrino oscillation as well as the scattering with background neutrinos,
even if it is not at the onset of the production.
Thus, it is currently consistent with the IceCube results.
Some deviation from the flavor-universality may be helpful 
to exclude the early-decay scenario in the future.

Finally, we comment on the multi-messenger approach.
It is a powerful way to explore the property of the source to combine the IceCube data
with the gamma-ray observation such as the Fermi collaboration~\cite{Ackermann:2014usa}.
In fact, regardless of production mechanism $pp$ or $p\gamma$, 
the authors in Refs.~\cite{Murase:2013rfa, Murase:2015xka} 
argued that the astrophysical source should be opaque to 
$1\mathchar`- 100\,\text{GeV}$ gamma-rays if it is the origin of the IceCube neutrino events.
The same method is also useful in the case of the DM decay scenario.
In Ref.~\cite{Murase:2015gea}, it is discussed that although the DM decay
scenario is currently consistent with the gamma-ray observation, it will be tested
by near-future gamma-ray observations.
In contrast to these scenarios, in the early-decay scenario, photons produced by $X$
are absorbed into the CMB since they are produced in the early epoch.
Thus, the present gamma-ray observations are not affected by the early-decay scenario.
Instead, as we discussed before, these photons might cause the CMB spectral distortions.

\section{Phenomenology}
\label{sec:pheno}
\setcounter{equation}{0}

\subsection{Production mechanism}
\label{subsec:prod}
As we saw in Sec.~\ref{subsec:present_flux}, the desired yield value of $X$ is 
\begin{align}
Y_{X} \sim 10^{-26},
\label{eq:yield}
\end{align}
to explain the neutrino flux observed by the IceCube experiment.
From now we discuss how such a tiny amount of $X$ is produced in the early universe.
Our purpose here is to show one specific example,
and hence the discussion is not generic.

\vspace{0.25cm}

We first give an overview of our production scenario.
In this paper, we concentrate on the case where
the inflaton mass $m_\phi$ is larger than $m_X$.
To be more specific, we take the inflaton mass to be
$m_\phi \sim \mathcal{O}(10^{11}\,\mathrm{GeV})$,
which is the case for, \textit{e.g.}, some class of new inflation 
models~\cite{Kumekawa:1994gx,Takahashi:2013cxa}.\footnote{
	As we will see later, this type of models is also compatible with the axion dark matter scenario.
} Then, there are four ways to produce $X$ after inflation:

\begin{enumerate}
\renewcommand{\labelenumi}{(\arabic{enumi})}

\item Inflaton decay

The inflaton must decay to transfer its energy density to ordinary matters.
If the branching ratio of the inflaton $\phi$ decaying into $X$ is nonzero, $X$ can be produced
from the inflaton decay.

\item Thermal production by dilute plasma

There exists radiation called dilute plasma even before the reheating completes.
The maximal temperature of the dilute plasma $T_\mathrm{max}$ is typically much higher than
the reheating temperature $T_R$, and hence it may thermally produce
the massive particle $X$.

\item Non-thermal production through inelastic scattering

Decay products of the inflaton with its energy $\sim m_\phi$ may produce
$X$ through inelastic scattering with the dilute plasma if 
$T_\mathrm{max} \gg m_X^2/m_\phi$~\cite{Harigaya:2014waa}.

\item Scattering among inflaton decay products before thermalization

High-energy particles with its energy $\sim m_\phi$ dominates the
energy density of the radiation sector before the thermalization of the dilute plasma 
completes~\cite{Kurkela:2011ti,Harigaya:2013vwa}.
Scatterings of such high-energy particles among themselves can also produce $X$.

\end{enumerate}
Since the desired yield value~\eqref{eq:yield} is tiny,
$X$ is typically overproduced if it is once
in thermal equilibrium in the early universe.
Even the non-thermal production by the radiation tends to overproduce $X$.
Thus, we take the reheating temperature $T_R$ to be low, of order $\mathcal{O}(10^{3}\,\text{GeV})$,
so that the production mechanisms~(2) and~(3) can be neglected.
As we will see below, the contribution from the production mechanism~(4)
is also negligible in such a case.
Instead, we assume that $X$ is mainly produced via a many-body decay process of $\phi$.
In such a case, the branching ratio of $\phi$ decaying to $X$ is well-suppressed by couplings and the phase space factor,
and hence we can naturally control the yield of $X$ produced by
the mechanism~(1).

\vspace{0.25cm}

Now we explain details of the production mechanism.
We consider the inflaton $\phi$ which couples only with the right-handed neutrino via the 
following interaction:
\begin{align}
\mathcal{L}_{\phi\nu\nu} = -\frac{1}{2}\lambda \phi \overline{\nu_{R}^{c}}\nu_{R} + \mathrm{h.c.},
\label{eq:scalar_decay}
\end{align}
where $\nu_{R}$ is the right-handed neutrino and $\nu_R^c = \mathcal{C} \bar{\nu}_R^{T}$
where the charge conjugation matrix $\mathcal{C}$ is given as 
$\mathcal{C} = i\gamma_0\gamma_2$ in the Weyl representation.
We denote the masses of $\phi$ and $\nu_{R}$ as $m_{\phi}$ and $m_{N}$, respectively.
The right-handed neutrino couples to the SM Higgs and leptons as usual:
\begin{align}
\mathcal{L}_{y} = - y \bar{L}\tilde{H}\nu_{R} + \mathrm{h.c.},
\end{align}
where $L$ is the SM left-handed lepton doublet, 
$\tilde{H} = \left(\epsilon H\right)^{\dagger}$ with $\epsilon$ being the antisymmetric
tensor of SU(2), and $y$ is the matrix of the Yukawa coupling.
Here we have suppressed the isospin and the generation indices for simplicity.
Note that some components of $y$ can be arbitrarily small even assuming 
the see-saw mechanism~\cite{GellMann:1980vs, Yanagida:1979as, Mohapatra:1979ia,Schechter:1980gr,Schechter:1981cv}.
This is because only the mass squared differences of the left-handed neutrinos are currently observed.

\begin{figure}[t]
\begin{minipage}{0.5\hsize}
\begin{center}
\includegraphics[scale = 0.8]{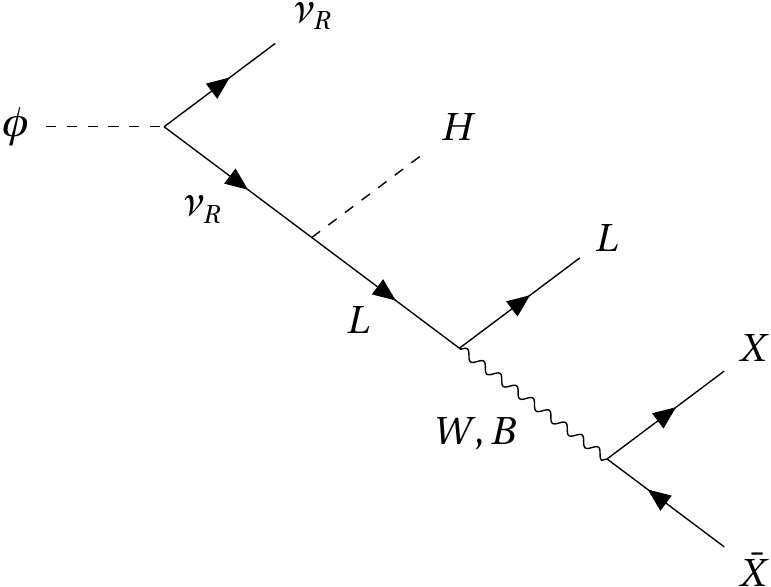}
\end{center}
\end{minipage}
\begin{minipage}{0.5\hsize}
\begin{center}
\includegraphics[scale = 0.8]{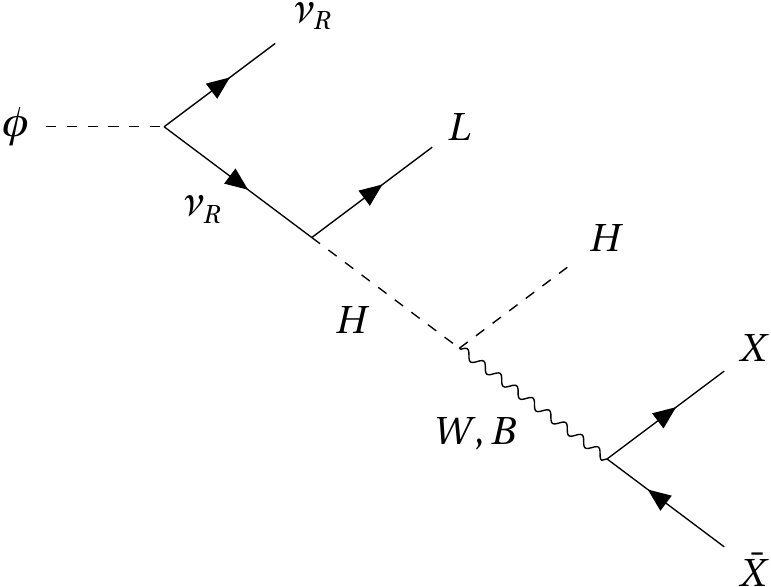}
\end{center}
\end{minipage}
\caption { \small
The Feynman diagrams for the five-body decay of $\phi$ to $X$.
The $\mathrm{SU}(2)_L$ and $\mathrm{U}(1)_Y$ gauge bosons are denoted
as $W$ and $B$, respectively. Note that the electroweak symmetry is restored
at the reheating epoch, and hence we do not distinguish 
each component of the $\mathrm{SU}(2)_L$ doublets.
They are generated using {\tt TikZ-Feynman}~\cite{Ellis:2016jkw}.
}
\label{fig:feyn_diag}
\end{figure}

The inflaton starts to oscillate around the bottom of its potential after inflation,
and the reheating proceeds through the interaction~\eqref{eq:scalar_decay}.
The decay rate of $\phi$ is estimated as
\begin{align}
\Gamma_{\phi} \simeq \mathcal{O}(0.01)\,\lambda^{2}m_{\phi},
\end{align}
and the reheating temperature is given by
\begin{align}
T_{R} &\sim \sqrt{M_{P}\Gamma_{\phi}} \nonumber \\
&\sim 10^{3}\,{\rm GeV} \left(\frac{\lambda}{10^{-11}}\right)\left(\frac{m_{\phi}}{10^{11}\,{\rm GeV}}\right)^{1/2}.
\label{eq:temp_rh}
\end{align}
We assume that the inequalities
\begin{align}
m_\phi ~ > ~ m_{X} \sim \mathcal{O}(10^{10}\,{\rm GeV}) ~ > ~ m_N,
\end{align}
are satisfied.\footnote{
	Since $\lambda$ is tiny [see Eq.~\eqref{eq:temp_rh}], 
	the right-handed neutrino mass induced by 
	the interaction~\eqref{eq:scalar_decay} is
	small compared to $m_X$ even if $\phi \sim M_P$.
	Note that $\phi \ll M_P$ typically holds for new inflation models.
} Then, $\phi$ directly decays into $X$ through five-body decay processes
$\phi \rightarrow X\bar{X}\nu_{R}LH$
[see Fig.~\ref{fig:feyn_diag}] 
since $X$ couples to the Standard Model particles via the gauge interaction.\footnote{
The SU(2) gauge symmetry is usually restored at the time of 
the inflaton decay, and hence we do not distinguish 
each component of the SU(2) doublets here.
}
The five-body decay rate is estimated as
\begin{align}
\Gamma_\text{five} \simeq \frac{c \lambda^2 y^2 g^4}{(4\pi)^{7}}m_{\phi},
\end{align}
where $c$ is a numerical factor and $g$ is the SU(2) or U(1) gauge coupling constant. 
The factor $(4\pi)^{-7}$ comes from the phase-space suppression of the five-body decay.
We have verified $c \lesssim \mathcal{O}(0.1)$ in the case of our interest 
by using {\tt FeynRules}~\cite{Alloul:2013bka} and {\tt MadGraph5}~\cite{Alwall:2014hca}.\footnote{
If $m_{\phi}$ is close to $m_{X}$, the numerical factor $c$ can be much smaller than unity.
}
Note that only the off-shell processes contribute due to the inequality $m_{N} < m_{X}$.
The yield of $X$ produced via this process is estimated as
\begin{align}
	Y_{X}^{(1)}
	&= \left[\frac{n_{X}}{s}\right]_{T=T_{R}} \sim \frac{T_{R}}{m_{\phi}}\frac{\Gamma_\text{five}}{\Gamma_{\phi}} \nonumber \\
	&\simeq 10^{-26}\left(\frac{c}{0.1}\right)
	\left(\frac{y^2}{10^{-12}}\right)
	\left(\frac{m_{\phi}}{10^{11}\,{\rm GeV}}\right)^{-1}
	\left(\frac{T_{R}}{10^{3}\,{\rm GeV}}\right).
\end{align}
Thus, the tiny amount of $X$ can be explained via the mechanism~(1)
for reasonable model parameters.

Next, we consider the production mechanism~(2) and~(3). 
It is well-known that there exists a dilute plasma even before the reheating completes.
If $m_{\phi} \gg T_{R}$, the particles produced by $\phi$ are initially under-occupied 
compared with the thermal distribution. 
In such a case, these particles do not instantaneously thermalize.
Instead, the so-called ``bottom-up thermalization" process occurs~\cite{Kurkela:2011ti,Harigaya:2013vwa}.
The maximal temperature of such a dilute plasma is estimated as
\begin{align}
T_{\rm max} &\sim \left(\frac{\alpha_{s}^4M_{P}^2}{T_{R}m_\phi}\right)^{1/5}T_{R} \nonumber \\
&\sim 10^7\, {\rm GeV}\, \left(\frac{\alpha_{s}}{0.1}\right)^{4/5}\left(\frac{T_{R}}{10^3\, {\rm GeV}}\right)^{4/5}\left(\frac{10^{11}\,{\rm GeV}}{m_{\phi}}\right)^{1/5},
\end{align}
where $\alpha_s$ is the fine structure constant of the SU(3) gauge group.
If $T_\mathrm{max}$ approaches to $m_X$, 
the production mechanisms~(2) and~(3) typically overproduce $X$.
Thus, we limit ourselves to the case with low $T_R$,
\textit{e.g.} of order $\mathcal{O}(10^{3}\,\text{GeV})$.
Then, the thermal production~(2) is well suppressed by
the Boltzmann factor.
The non-thermal production~(3) is also negligible
since $T_{\rm max} < m_{X}^2/2m_{\phi}$ in this case.

Now let us move on to the production mechanism~(4).
Since we take the reheating temperature to be low,
it takes a long time for the dilute plasma to achieve the thermalization.
Before the thermalization is achieved, 
the energy density of the decay products is dominated by high-energy
particles whose typical energy is of order $\sim m_\phi$~\cite{Harigaya:2013vwa}.
These particles can produce $X$ by scattering with each other.
The number density of high-energy particles is estimated as
\begin{align}
	n_{\rm high} 
	\sim \frac{\rho_\phi \Gamma_\phi}{m_\phi H} 
	\sim \frac{M_P H T_R^2}{m_\phi},
\end{align}
and hence the number density of $X$ produced 
during one Hubble time is given by
\begin{align}
	n_X &\sim \frac{\langle \sigma v\rangle n_{\rm high}^2}{H} \nonumber \\
	&\sim \frac{\alpha_s^2M_P^2 H T_R^4}{m_\phi^4},
\end{align}
where we have estimated the averaged cross section as $\langle \sigma v\rangle \sim \alpha_s^2/m_\phi^2$.
The number density of $X$ produced at $H = H_p$ contributes to the yield as
\begin{align}
	Y_X^{(4)}
	&= \left[\frac{n_X}{s} \right]_{T_R} 
	\sim \left.\frac{T_R}{m_\phi}\frac{n_X}{n_\phi}\right\rvert_{T_R}
	\sim \left.\frac{T_R}{m_\phi}\frac{n_X}{n_\phi}\right\rvert_{H = H_p}
	\sim \alpha_s^2 \left(\frac{T_R}{m_\phi}\right)^4\left(\frac{T_R}{H_p}\right).
\end{align}
Thus, the late time production gives the dominant contribution.
Just before the thermalization, the Hubble parameter is estimated as
\begin{align}
	H(T=T_\mathrm{max}) 
	\sim \frac{T_\mathrm{max}^4}{M_P T_R^2} \sim \left(\frac{\alpha_s^4 M_P^2}{T_R m_\phi}\right)^{4/5}\frac{T_R^2}{M_P},
\end{align}
and hence the yield of $X$ produced by the mechanism~(4) is given by
\begin{align}
	Y_X^{(4)} 
	&\sim \alpha_s^{-6/5}\left(\frac{T_R}{M_P}\right)^{3/5}\left(\frac{T_R}{m_\phi}\right)^{16/5} \nonumber \\
	&\sim 10^{-34} \left(\frac{0.1}{\alpha_s}\right)^{6/5}
	\left(\frac{T_R}{10^{3}\,\mathrm{GeV}}\right)^{19/5}
	\left(\frac{10^{11}\,\mathrm{GeV}}{m_\phi}\right)^{16/5}.
\end{align}
Therefore, it is negligible for 
$T_R \sim \mathcal{O}(10^3\,\mathrm{GeV})$ and $m_\phi \sim \mathcal{O}(10^{11}\,\mathrm{GeV})$.

Finally, we comment on the baryogenesis.  In our scenario, the reheating
temperature should be low enough to avoid the overproduction of $X$.
In such a case, the thermal \cite{Fukugita:1986hr} and non-thermal
\cite{Kumekawa:1994gx,Asaka:1999yd,Asaka:1999jb} leptogenesis
scenarios do not work in its original form.  Still, if the heavy
neutrino masses are sufficiently degenerate, the leptogenesis may be
possible~\cite{Pilaftsis:2003gt}.
The Affleck-Dine~\cite{Affleck:1984fy} or electroweak
baryogenesis~\cite{Kuzmin:1985mm,Shaposhnikov:1986jp,
  Shaposhnikov:1987tw} may be other possibilities to explain the
baryon to photon ratio of the universe.

\subsection{Dark matter}
\label{subsec:dm}
In this subsection, we show that the our scenario naturally accommodates the axion dark matter (DM).
In order to see this point, let us consider the dynamics of the PQ symmetry breaking field in the early universe.

We assume that the inflation scale is low, 
such as those of the models in Refs.~\cite{Kumekawa:1994gx,Takahashi:2013cxa}.
Note that the mass scale of the inflaton is $m_\phi \sim \mathcal{O}(10^{11}\,\text{GeV})$
in these models, and hence it is consistent with our production scenario.
Then, the PQ symmetry is already broken during inflation.
In such a case, the coherent oscillation of the axion provides 
a natural candidate of DM.\footnote{
For a review on the axion cosmology, see, \textit{e.g.} Ref.~\cite{Kawasaki:2013ae}.
}
The present energy density of the axion is given by~\cite{Turner:1985si}
\begin{align}
\Omega_{a}h^{2} = 0.18\,\theta_{a}^{2}
\left(\frac{f_{a}}{10^{12}\,{\rm GeV}}\right)^{1.19}\left(\frac{\Lambda_\text{QCD}}{400\,{\rm MeV}}\right),
\end{align}
where $\theta_{a}$ is the miss-alignment angle at the onset of the coherent oscillation and 
$\Lambda_\text{QCD}$ is the energy scale of the QCD phase transition.
For $f_{a} \sim \mathcal{O}(10^{11}\mathchar`-10^{12}\,{\rm GeV})$, the axion coherent oscillation can
provide the present DM abundance if $\theta_{a} \sim \mathcal{O}(1)$.
Thus, our scenario is consistent with the axion DM scenario.

Here we comment on the isocurvature perturbation. 
If the PQ symmetry is broken during inflation, 
the axion already exists and acquires fluctuations during that epoch.
They contribute to the DM isocurvature perturbation, which is constrained by the observation~\cite{Ade:2015xua}.
In our case, however, the inflation scale is taken to be low, and thus the isocurvature 
perturbation is suppressed by the factor $H_\text{inf}/f_a$ where $H_\text{inf}$ is the Hubble parameter
during inflation. It is consistent with the observation for
$H_\text{inf}/f_a \lesssim 10^{-6}$.

\subsection{Lifetimes of other particles}
\label{subsec:lifetime}
We consider the neutral component of a heavy SU(2) doublet as $X$, which is a mother particle
of the high energy neutrinos observed by IceCube. There are also the charged component 
of the SU(2) doublet and a heavy SU(3) triplet in the theory. 
In this subsection, we discuss fates of these particles
in the early universe.

First, let us consider the heavy SU(3) triplet (denoted as $X_{3}$). 
We assume that they are more massive than the SU(2)
doublet that includes $X$. Note that it is usually the case due to the RG effects. 
Since $X_3$ and $X_2$ are embedded into the same GUT multiplet, 
each component of $X_{3}$ can decay into $X_{2}$ 
by exchanging massive gauge bosons.
The decay rate is roughly estimated as
\begin{align}
\Gamma_{X_{3}} \sim \frac{g^{4}}{\left(4\pi\right)^3}\frac{m_{X_{3}}^5}{m_{V}^4},
\end{align}
where $g$ is the gauge coupling of the GUT, $m_{X_{3}}$ is the mass of $X_{3}$ and
$m_{V}$ is the mass of the massive gauge bosons
which is of the order of the GUT scale. 
If we take $m_{V} \sim \mathcal{O}(10^{16}\,{\rm GeV})$ and 
$m_{X_3} \sim \mathcal{O}(10^{11}\,{\rm GeV})$, for example,
the decay rate is estimated as $\Gamma_{X_{3}} \sim 10^{-7}\,{\rm GeV}$.
Thus, $X_{3}$ does not affect any observables because it decays 
well before the BBN.
Moreover, the yield of $X_{3}$ produced in the early universe is at most comparable to that of $X_{2}$, and 
hence the estimation of $Y_{X}$ in the previous subsection still holds.

Next, let us consider the charged component of $X_{2}$. We denote it as $X^{\pm}$ here. Before the electroweak
symmetry breaking, it is stable because the mass of $X^{\pm}$ is exactly the same as that of $X$.
However, a mass difference is generated radiatively after the electroweak symmetry breaking. In the limit $m_{X} \gg m_{Z}$
where $m_{Z}$ is the $Z$-boson mass, the mass difference is given by~\cite{Thomas:1998wy, Nagata:2014wma}
\begin{align}
\delta m \equiv m_{X^{\pm}} - m_{X} \simeq \frac{1}{2}\alpha_{2}\sin^{2}\theta_{W}m_{Z}
\simeq 350\,\text{MeV},
\end{align}
where $\alpha_2$ is the fine structure constant of the SU(2) gauge group, $\theta_{W}$ is the weak mixing
angle and $m_{X^{\pm}}$ is the mass of $X^{\pm}$. 
Due to this mass difference, $X^{\pm}$ decays into $X$
by mainly emitting a soft charged pion~\cite{Thomas:1998wy, Nagata:2014wma}. 
The lifetime is of order $\mathcal{O}(10^{-10}\,{\rm sec})$,
and hence it starts to decay just after the electroweak phase transition, 
well before the BBN. Thus, it has no effect on any observables.
Again, the estimation of $Y_{X}$ is not affected by this process 
since the yield of $X^{\pm}$ is at most comparable to $Y_{X}$.

\section{Summary}
\label{sec:summary}
IceCube detected high-energy neutrino events whose origin is still unknown.
There have been a lot of discussion on the origin from both astrophysical and
particle physics point of view.
As one possibility, in Refs.~\cite{Ema:2013nda,Ema:2014ufa}, 
it was proposed that decay of 
a massive long-lived particle in the early universe
can be the origin of the IceCube neutrino events.
We have called such a scenario as 
``early decay'' scenario, and the massive long-lived particle
as $X$ in this paper. Interestingly, 
it was shown that the mass scale of $X$ can be 
as large as of order $\mathcal{O}(10^{10}\,\text{GeV})$
if it decays well before the present universe,
although the IceCube neutrino events are in the energy range of 
$\mathcal{O}(10\,\text{TeV}) \lesssim E_\nu \lesssim \mathcal{O}(1\,\text{PeV})$.

In this paper, we have extended the discussion in Refs.~\cite{Ema:2013nda,Ema:2014ufa}.
We have constructed a specific particle physics model 
that includes $X$ based on the Peccei-Quinn model.
The hadronic axion models include an additional heavy quark,
and there is also an SU(2) counterpart once we embed the model into the grand unified theory.
We have identified the charge neutral component of 
the SU(2) doublet as $X$.
We have calculated the energy spectrum of neutrinos produced from
the decay of $X$, and followed the time evolution of the neutrino flux 
including the scattering processes with the background neutrinos.
The decay of $X$ also produces electromagnetic particles such as photons,
electrons and positrons. They induce electromagnetic cascades 
with the background photons, and
secondary neutrinos produced in such cascade processes are also 
included in our calculation.
We have shown that our model can explain the IceCube neutrino events 
in the energy region of 
$\mathcal{O}(10\,\text{TeV}) \lesssim E_\nu \lesssim \mathcal{O}(1\,\text{PeV})$.
The favored mass of $X$ is of order $m_X \sim \mathcal{O}(10^{10}\,\text{GeV})$,
and the favored lifetime is of order $\tau_X \sim \mathcal{O}(10^{11}\,\text{sec})$.
The decay of $X$ can also slightly modify the thermal history of the universe.
We have checked that our scenario is currently consistent with the CMB and BBN constraints.  
Notice that our scenario predicts the $y$ and $\mu$ parameters to be 
$y \sim \mathcal{O}(10^{-9})$ and $\mu \sim \mathcal{O}(10^{10})$,
respectively.  Thus, future CMB observations such as PIXIE~\cite{Kogut:2011xw} and
PRISM~\cite{Andre:2013nfa} have a possibility to detect the CMB distortions caused by the
present scenario.

We have also discussed the cosmological history based on our model.
In particular, the yield of $X$ should be of order $\mathcal{O}(10^{-26})$
to reproduce the normalization of the IceCube neutrino flux.
This is a rather small value, and hence we have discussed in detail 
how such a tiny amount of $X$ is produced in the early universe.
The thermal as well as non-thermal production by background radiation 
typically overproduce $X$, and hence we limit ourselves 
to the case of the low reheating temperature.
Instead, we have used a many-body decay of the inflaton
as a production mechanism. In such a case, the phase-space factor as well as
couplings suppress the abundance of $X$.
The dark matter abundance and fate of additional particles
other than $X$ are also briefly discussed.

\section*{Acknowledgments}
The work of T.M. was supported by Grant-in-Aid for Scientific research No. 26400239.
The work of Y.E. was supported in part by JSPS Research Fellowships for Young Scientists 
and the Program for Leading Graduate Schools, MEXT, Japan.

\appendix

\small
\bibliography{ref}

\end{document}